\documentclass[twocolumn,twoside]{article}
\usepackage{epsfig}
\usepackage{latexsym}
\usepackage{amsmath}
\usepackage{graphics}
\def\br{{\bf r}}

\def\bv{{\bf v}}

\def\d{{\rm d}}

\begin{document}

\title{\Large\bf Exact hydrodynamics of a trapped dipolar
Bose-Einstein condensate}
\author{{\large Duncan H J O'Dell${}^\sharp$, Stefano Giovanazzi${}^\flat$,
Claudia Eberlein${}^\sharp$}\\
        {${}^\sharp$\normalsize\em Dept of Physics \& Astronomy,
    University of Sussex,}\\
    {\normalsize\em Falmer, Brighton BN1 9QH, England}\\
    {${}^\flat$\normalsize\em School of Physics \& Astronomy,
    University of St Andrews,}\\
    {\normalsize\em North Haugh, St Andrews KY16 9SS, Scotland}\\
    {\normalsize (Submitted \today)}}
\date{\parbox{140mm}{\small
\hspace*{3mm} We derive the exact density profile of a
harmonically trapped Bose-Einstein condensate (BEC) which has
dipole-dipole interactions as well as the usual $s$-wave contact
interaction, in the Thomas-Fermi limit. Remarkably, despite the
non-local anisotropic nature of the dipolar interaction, the
density turns out to be an inverted parabola, just as in the pure
$s$-wave case, but with a modified aspect ratio. The ``scaling''
solution approach of Kagan, Surkov, and Shlyapnikov [Phys. Rev. A
{\bf 54}, 1753 (1996)] and Castin and Dum [Phys. Rev. Lett. {\bf
77}, 5315 (1996)] for a BEC in a time-dependent trap can therefore
be applied to a dipolar BEC, and we use it to obtain the exact
monopole and quadrupole shape oscillation frequencies.
}} \maketitle

\noindent The current revolution in the cooling of atomic gases to
ultra-cold temperatures has led to the experimental realization of
a number of highly idealized and fundamental regimes in the
quantum physics of many-particle systems
\cite{ketterle_varenna,dalfovo99}. For example, in a Bose-Einstein
condensed gas containing several million atoms the zero-point
kinetic energy can be negligible in comparison to the
interparticle interaction energy. In this so-called Thomas-Fermi
regime the collective dynamics of the Bose-Einstein condensate
(BEC) are described by the collisionless hydrodynamic theory of
Bose superfluids at zero temperature \cite{stringari96}.
Furthermore, the interparticle interactions can be directly
manipulated using external electromagnetic fields, giving a degree
of freedom not normally available in the systems traditionally
studied in statistical physics. The dominant interactions in these
gases are usually of the van der Waals type, decaying as $r^{-6}$,
short-ranged in comparison to the interatomic separation. This
permits an effective description of the interactions solely in
terms of the $s$-wave scattering length, $a_{\mathrm{s}}$. Within
the mean-field scheme of the Gross-Pitaevskii equation
\cite{gp-eqn} the interactions are modelled by the contact
pseudo-potential, $g \delta(\br) \equiv (4 \pi a_{\mathrm{s}}
\hbar^{2}/m) \delta(\br)$,
 where $m$
is the atomic mass, and $g$ incorporates the quantum aspects of
low-energy scattering. By using a magnetic field to induce a
Feshbach resonance during atomic collisions, one can adjust the
value of $a_{\mathrm{s}}$ between positive (repulsive) and
negative (attractive) values \cite{inouye}.

Here we consider a harmonically trapped BEC which, as well as van
der Waals interactions, also has dipole-dipole interactions. Such
a system was first considered by Yi and You \cite{yi00} in the
context of dipoles induced by an external electric field and by
G{\'{o}}ral, Rz{\c{a}}{\.{z}}ewski, and Pfau \cite{goral00} in the
context of permanent magnetic dipoles aligned by an external
magnetic field. In comparison to van der Waals interactions,
dipole-dipole interactions have a much longer range and are
anisotropic. This has been predicted to lead to BECs with unusual
stability properties \cite{santos00,lushnikov}, exotic ground
states such as supersolid \cite{giovanazzi2002a,goral2002} and
checkerboard phases \cite{goral2002}, and modified excitation
spectra \cite{yi01,goral2002b} even to the extent of a roton
minimum \cite{odell03}. Dipole-dipole interactions are also
inherently controllable, either via the magnitude of the external
electric field in the case of electrically induced dipoles, or in
the case of permanent dipoles by modulating the external aligning
field in time \cite{giovanazzi2002b}, which allows control over
both the magnitude and sign of the dipolar interactions.

The long-range part of the interaction between two dipoles
separated by $\br$, and aligned by an external field along a unit
vector $\hat{\mathbf{e}}$, is given by
\begin{equation}
U_{\mathrm{dd}}(\br)= \frac{C_{\mathrm{dd}}}{4 \pi}\, \hat{{\rm
e}}_{i} \hat{{\rm e}}_{j} \frac{\left(\delta_{i j}- 3 \hat{r}_{i}
\hat{r}_{j}\right)}{r^{3}} \label{eq:staticdipdip}
\end{equation}
where the coupling $C_{\mathrm{dd}}/ (4 \pi)$ depends on the
specific realization. Electric dipoles induced by an external
static electric field $\mathbf{E}=E\hat{\mathbf{e}}$  have $
C_{\mathrm{dd}}=E^{2} \alpha^{2}/\epsilon_{0}$  \cite{yi00}, where
$\alpha$ is the static polarizability. For magnetic atoms with
magnetic dipole moment $d_{m}$ aligned by an external magnetic
field $\mathbf{B}=B\hat{\mathbf{e}}$, then
$C_{\mathrm{dd}}=\mu_{0} d_{m}^{2}$. A measure of the strength of
the dipole-dipole interaction relative to the short-range van der
Waals $s$-wave scattering energy is given by the dimensionless
quantity
\begin{eqnarray}
\varepsilon_{\mathrm{dd}}  \equiv \frac{C_{\mathrm{dd}}}{3 g} \ .
\label{epsilon}
\end{eqnarray}
For $\varepsilon_{\mathrm{dd}}
>1$ instabilities are expected \cite{goral00,giovanazzi2002b} in the
limit of negligible kinetic energy. Note that the value of $g$ can
be altered in the presence of a strong external electric field
\cite{yi01}. Alkali atoms can have a magnetic dipole moment of $
d_{m} = 1 \mu_{\mathrm{B}}$ (Bohr magneton), and for $^{87}$Rb,
which has $a_{\mathrm{s}} \approx 103 \mathrm{a}_{0}$ (Bohr
radius), one finds $\varepsilon_{\mathrm{dd}} \approx 0.007$. Na
has $\varepsilon_{\mathrm{dd}} \approx 0.004$, and accordingly
magnetic dipolar effects in BECs of these atoms are expected to be
small, at least in a stationary condition. The effects of the
dipolar interactions can be made visible by rotating the magnetic
field in resonance with a collective excitation frequency of the
system \cite{giovanazzi2002b}. Also, using a Feshbach resonance to
reduce $a_{\mathrm{s}}$ could substantially increase
$\varepsilon_{\mathrm{dd}}$ \cite{yi03}. However, a good candidate
atom for a dipolar BEC is chromium which has $d_{m}=6
\mu_{\mathrm{B}}$. The $s$-wave scattering length of $^{52}$Cr,
the more common isotope, is $170 \pm 40$ a$_0$. Consequently for
that isotope $\varepsilon_{\mathrm{dd}} \approx 0.089$.  The less
common isotope $^{50}$Cr has $a_{\mathrm{s}} = 40 \pm 15$ a$_0$,
and $\varepsilon_{\mathrm{dd}} \approx 0.36$ is much higher. A
chromium BEC does not yet exist, but a program of experiments is
underway with that goal in mind \cite{pfau}. Using a crossed
optical trap, like in the recent cesium condensation experiment
\cite{weber03}, atom numbers in a $^{52}$Cr BEC on the order of
$N= 10^4$ might be within reach, and the radial and longitudinal
frequencies of the cylindrically symmetric (optical) harmonic
confinement would be in the region of $\omega_x=\omega_y= 2 \pi
170$ s$^{-1}$ and $\omega_z= 2 \pi 240$ s$^{-1}$, respectively
\cite{pfau}. The trap anisotropy $\gamma=\omega_z/\omega_x=1.41$
is optimal for enhancing the condensate shape deformations
\cite{giovanazzi03}. For a harmonically trapped BEC with just
$s$-wave scattering the Thomas-Fermi limit is reached for large
values of the dimensionless parameter $N
a_{\mathrm{s}}/a_{\mathrm{ho}}$ \cite{dalfovo99} where
$a_{\mathrm{ho}}=\sqrt{\hbar/m \omega}$ is the harmonic oscillator
length of the trap, and $N$ is the total number of atoms. For
$^{52}$Cr one has $Na_{\mathrm{s}}/a_{\mathrm{ho}}\approx 90$,
illustrating that the Thomas-Fermi regime is an appropriate
starting point for analyzing these experiments.

The application of collisionless hydrodynamics to an ordinary BEC
is based upon the existence of a macroscopic condensate order
parameter $\psi(\br,t)=\sqrt{n(\br,t)} \exp \mathrm{i}
\phi(\br,t)$, normalized to $N$, consisting of density $n(\br,t)$,
and phase $\phi(\br,t)$. The phase determines the superfluid
velocity via potential flow, $\mathbf{v}(\br,t)=(\hbar/m)
\mathbf{\nabla} \phi(\br,t)$. The nonlinear evolution of the
condensate in the Thomas-Fermi regime is then dictated by the
continuity and Euler equations given by, respectively,
\begin{eqnarray}
{\partial n \over \partial t} &=& - {\bf \nabla} \cdot (n \bv) \;,\label{continuity}\\
m{\partial {\bf v} \over \partial t} &=&- {\bf \nabla} \left(
\frac { m v^2}{2} + V_{\mathrm{ext}} + g n\right) \;.\label{euler}
\end{eqnarray}
$V_{\mathrm{ext}}$ is the external potential, which is provided by
the harmonic trap, $V_{\mathrm{ext}}({\bf r})= m( \omega_x^2 x^2+
\omega_y^2 y^2 +  \omega_z^2 z^2)/2$. Eqs.\  (\ref{continuity})
and (\ref{euler}) describe the potential flow of a fluid whose
pressure and density are related by the equation of state $P
=(g/2) n^2$,  and are equivalent to the time dependent
Gross-Pitaevskii equation \cite{gp-eqn} when the zero-point
fluctuation term $(-\hbar^2 \nabla^2 \sqrt{n} / 2 m \sqrt{n})$ is
included in the brackets on the right hand side of (\ref{euler})
after the Bernoulli term $m v^2 / 2$. The equilibrium solution of
(\ref{continuity}) and (\ref{euler}), which has $\bv=0$, is the
well known Thomas-Fermi inverted parabola density profile
\begin{equation}
n(\br)=[\mu-V_{\mathrm{ext}}(\br)]/g \ \ \mbox{ for }\
n(\br)\geq0,
\end{equation}
and $n(\br)=0$ elsewhere. Thus the density profile is completely
determined by the trapping potential, having the same aspect ratio
as the trap.

An analysis of the dynamical ($\bv \neq 0$) solutions of
(\ref{continuity}) and (\ref{euler}) has been given by Stringari
\cite{stringari96}. However, motivated by the experimental
possibility of changing the trapping frequencies in time, so that
$\omega_{j} \rightarrow \omega_{j}(t)$ with $j=x,y,z$, Kagan,
Surkov, and Shlyapnikov \cite{kagan} and Castin and Dum
\cite{castin} studied a special class of ``\textbf{scaling}''
solution, valid for time-dependent harmonic traps, corresponding
to \cite{dalfovo99}
\begin{eqnarray}
n(\br,t) &=& n_0(t)
\left[1-\frac{x^2}{R_x^2(t)}-\frac{y^2}{R_y^2(t)}
 - \frac{z^2}{R_z^2(t)}\right]\;, \label{parabola} \\
\bv(\br,t) & = & \frac{1}{2} \mathbf{\nabla} \left[ \alpha_x(t)
x^{2} + \alpha_y(t) y^{2}+ \alpha_z(t) z^{2} \right]
\label{velocity}
\end{eqnarray}
valid where $n(\br,t) \geq0$ and $n(\br,t)=0$ elsewhere. The
central density $n_0(t)$ is constrained by normalization to be
$n_0(t) = 15 N / [8\pi  R_x(t) R_y(t) R_z(t)]$. The time-evolution
of the radii $R_j$ reduces to the solution of three
\emph{ordinary} differential equations, and the components of the
velocity field, $\alpha_{j}$, depend on the corresponding radii as
$\alpha_{j}= \dot{R}_{j}/R_{j}$. Although these scaling solutions
represent only a limited class of the possible solutions to the
hydrodynamic equations, they model two crucial experimental
situations: \textbf{i)} the expansion of a BEC when the trap is
relaxed or turned off---this is usually necessary to image the BEC
which would otherwise be too small \textit{in situ} (time of
flight measurements then record the original momentum
distribution); \textbf{ii)} the response of the BEC to a
modulation of the trapping frequencies, which is the easiest way
to generate shape oscillations of the BEC.

The existence of the class (\ref{parabola}) of analytic scaling
solutions relies upon a harmonic trap and the local character of
$s$-wave interactions. The remainder of this paper demonstrates an
analogous exact result for a dipolar BEC. This result is not
obvious because of the non-local nature of dipolar interactions.
Nevertheless, it turns out that the dipolar contribution to the
mean-field potential generated by a parabolic density profile is
itself also parabolic so that (\ref{parabola}) and
(\ref{velocity}) still represent a self-consistent solution to the
hydrodynamic equations of motion.

\emph{Exact static solution for a Thomas-Fermi dipolar BEC}
--- Dipolar interactions can be included in a mean-field
treatment of the BEC via a Hartree potential
\begin{equation}
\Phi_{\mathrm{dd}}(\br) \equiv \int \d^{3}r'\
U_{\mathrm{dd}}(\br-\br') n(\br')
\end{equation}
as shown by Yi and You \cite{yi01} based on a calculation of the
two-body T-matrix. The hydrodynamic equations of motion for
dipolar BEC are then the same as above but with the Euler equation
(\ref{euler}) supplemented by adding $\Phi_{\mathrm{dd}}(\br)$
into the bracket on the r.h.s. We begin our analysis with a
suggestive re-casting of the dipole-dipole term
(\ref{eq:staticdipdip})
\begin{equation}
\frac{\left(\delta_{i j}- 3 \hat{r}_{i}
\hat{r}_{j}\right)}{4 \pi r^{3}}= - \nabla_{i}
\nabla_{j}\frac{1}{4 \pi r}-\frac{1}{3}
\delta_{ij} \delta(\br)\;.
\end{equation}
We can then write
\begin{eqnarray}
\Phi_{\mathrm{dd}}(\br) &&\hspace*{-6mm} =
-C_{\mathrm{dd}}\:\left( \hat{\mathrm{e}}_{i}
\hat{\mathrm{e}}_{j}  \nabla_{i} \nabla_{j}
\phi(\br)+\frac{1}{3}  n(\br)
\right) \label{eq:electrostaticgreen1}\\
\mbox{with }\ \phi(\br) &&\hspace*{-6mm} \equiv  \frac{1}{4 \pi}
\int \frac{\d^{3}r' \;n(\br')}{\vert \br-\br' \vert }\;.
\label{eq:electrostaticgreen2}
\end{eqnarray}
The problem thereby reduces to an analogy with electrostatics, and
one need only calculate the `potential' $\phi(\br)$ arising from
the `static charge' distribution $n(\br)$. We are now in a
position to see the basic form $n(\br)$ must take, since
$\phi(\br)$ given by (\ref{eq:electrostaticgreen2}) obeys
Poisson's equation $\nabla^{2} \phi= - n (\br)$. Thus, if the
density is parabolic, of the form (\ref{parabola}), then Poisson's
equation is satisfied by a potential of the form $
\phi=a_{0}+a_{1}x^{2}+a_{2}y^{2}+a_{3}z^{2}+a_{4}x^{2}y^{2}+a_{5}x^{2}z^{2}
 +a_{6}y^{2}z^{2}+a_{7}x^{4}+a_{8}y^{4}+a_{9}z^{4} $,
and by (\ref{eq:electrostaticgreen1}) $\Phi_{\mathrm{dd}}(\br)$ is
also parabolic. Therefore, an inverted parabola remains a
self-consistent solution to the dipolar hydrodynamic equations.
The evaluation of the integral (\ref{eq:electrostaticgreen2}) is
too long to be reproduced here \cite{future}, but can be achieved
either by integrating over successive thin ellipsoidal shells, or
in the case of cylindrical symmetry by transforming into
spheroidal coordinates and using the known Green's function for
Poisson's equation in that basis.

\begin{figure}[htbp]
\begin{center}
\centerline{\epsfig{figure=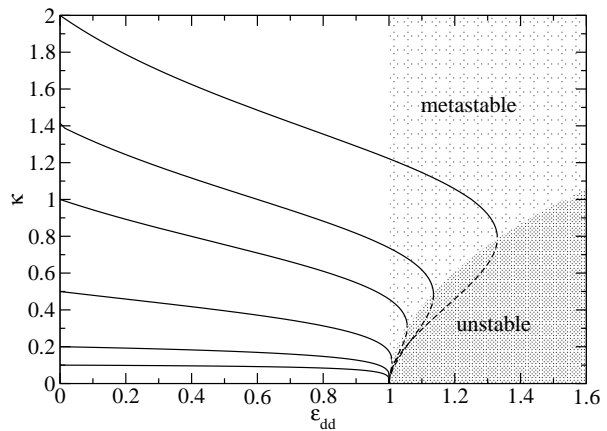,
  width= 7cm,angle=-90}}
\end{center}
\vspace{0ex} \caption{Equilibrium aspect ratio $\kappa$ of the
BEC, as a function of the dipole-dipole to $s$-wave coupling
ratio, $\varepsilon_{\mathrm{dd}}$. Each line is for a trap of a
different aspect ratio $\gamma$ ($\kappa=\gamma$ when
$\varepsilon_{\mathrm{dd}}=0$). When $0 < \gamma \ [ \kappa ] < 1$
the trap $[\mbox{condensate}]$ is prolate, $\gamma \ [ \kappa ]
> 1$ the trap $[\mbox{condensate}]$ is oblate.
 The dashed lines indicate an unstable
branch. However, local density perturbations most likely render
the condensate unstable whenever $\varepsilon_{\mathrm{dd}}>1$.}
\label{fig:aspect}
\end{figure}

For simplicity (though not necessity), we shall restrict attention
to cylindrical symmetry for both trap and BEC about the $z$ axis,
along which the external polarizing field is understood to point.
Thus we have $R_{x}=R_{y}$ for the BEC and $V_{\mathrm{ext}}=(m/2)
[ \omega_{x}^2\rho^{2} + \omega_{z}^{2}z^{2} ]$, where
$\rho^{2}=x^{2}+y^{2}$, for the trap. Then, inside the condensate
region, the dipolar potential (\ref{eq:electrostaticgreen1})
generated by an inverted parabola density distribution
(\ref{parabola}) of dipoles, is \cite{future}
\begin{eqnarray} \Phi_{\mathrm{dd}} =
\frac{ n_0 C_{\mathrm{dd}}}{3}\left[{\rho^2 \over R_{x}^2}-{2 z^2
\over R_{z}^2}-f \left( \kappa \right)\left(1-\frac{3}{2}{\rho^2 -
2 z^2 \over R_{x}^2- R_{z}^2}\right) \right]\label{phiddinside}
\\
\mbox{where} \quad
f(\kappa)=\frac{1+2\kappa^2}{1-\kappa^2}-\frac{3\kappa^2
\mathrm{arctanh}\sqrt{1-\kappa^2}}{\left(1-\kappa^2\right)^{3/2}}
\quad \label{f}
\end{eqnarray}
with $\kappa \equiv R_{x}/R_{z}$ being the condensate aspect
ratio. We recover the result (9) of \cite{giovanazzi2002b} in the
particular case of a spherical dipolar condensate ($R_x=R_z$). In
order to determine the equilibrium values of $R_{x}$ and $R_z$,
the result (\ref{phiddinside}) for $\Phi_{\mathrm{dd}}$ should be
substituted into the Euler equation (\ref{euler}) with $\bv=0$,
giving a transcendental equation for the aspect ratio $\kappa$
\cite{yi01}
\begin{equation}
3 \kappa^{2} \varepsilon_{\mathrm{dd}} \left[
\left(\frac{\gamma^{2}}{2}+1
\right)\frac{f(\kappa)}{1-\kappa^{2}}-1 \right] + \left(
\varepsilon_{\mathrm{dd}}-1 \right)
\left(\kappa^{2}-\gamma^{2}\right)  =0 \label{eq:transcendental}
\end{equation}
where $\gamma=\omega_{z}/\omega_{x}$ is the ratio of the harmonic
trapping frequencies. Fig.\ \ref{fig:aspect} shows some examples
of the dependence of $\kappa$ upon $\varepsilon_{\mathrm{dd}}$ for
oblate, spherical and prolate traps.
The effect of the dipole-dipole forces polarized along the
$z$-axis is to make the condensate more cigar-shaped along $z$.
For an oblate trap ($\gamma >1$) the BEC becomes exactly spherical
when $\varepsilon_{\mathrm{dd}}
=(5/2)(\gamma^{2}-1)/(\gamma^{2}+2)$. An analysis of the energy
functional associated with the scaling solution (\ref{parabola})
shows that when $\varepsilon_{\mathrm{dd}} > 1$ the solution loses
its global stability to \emph{scaling} perturbations and is only
metastable \cite{santos00,yi01,future} (only a local minimum of
the energy), the global minimum being a collapsed pencil-like
prolate state. Increasing $\varepsilon_{\mathrm{dd}}$ yet further,
the scaling solution ceases to exist entirely in the region marked
as unstable in Fig.\ \ref{fig:aspect}. However, we strongly
caution the reader that any state of a dipolar BEC with
$\varepsilon_{\mathrm{dd}}>1$ is probably unattainable since other
classes of perturbations (that do not respect the scaling form
(\ref{parabola})), such as local phonons, render the dipolar BEC
unstable to collapse whenever $\varepsilon_{\mathrm{dd}}>1$
\cite{goral00,giovanazzi2002b}.

A moment's thought shows that the aspect ratio is independent of
the specifics of the density profile and indeed the same
transcendental Eq.\ (\ref{eq:transcendental}) has been previously
obtained using a Gaussian ansatz \cite{yi01}. However, what is
specific to the exact solution are the absolute radii which, once
(\ref{eq:transcendental}) has been solved for $\kappa$, are given
by
\begin{eqnarray}
&&\displaystyle\hspace*{-10mm} R_{x}=R_{y}=\left[\frac{15 g N
\kappa}{4 \pi m \omega_{x}^{2}} \left\{1+ \displaystyle
\varepsilon_{\mathrm{dd}} \left( \frac{3}{2} \frac{
 \kappa^{2} f(\kappa)}{1-\kappa^{2}}-1 \right)
 \right\} \right]^{1/5} \label{eq:Rxsol}
\end{eqnarray}
and $R_{z}=R_{x}/\kappa$. If the trap is turned off the $s$-wave
and dipole-dipole interaction energies are converted into kinetic
energy, the so-called release energy $E_{\mathrm{rel}}$, which can
be directly measured in an experiment. One finds
$E_{\mathrm{rel}}=15 g
N^{2}[1-\varepsilon_{\mathrm{dd}}f(\kappa)]/(28 \pi R_{x}^{2}
R_{z})$.

\emph{Exact scaling hydrodynamics of a dipolar condensate} --- The
parabolic form of $\Phi_{\mathrm{dd}}(\br)$ means that the scaling
solutions \cite{kagan,castin} for dynamics also apply to a dipolar
BEC. Substituting the scaling solutions (\ref{parabola}) and
(\ref{velocity}) into the continuity and Euler equations yields
the following classical equations of motion for the radii
\begin{eqnarray}
\ddot{R}_{x} & = & - \omega_{x}^{2}(t)R_{x} +\frac{15 g N}{4 \pi m R_{x}
R_{z}}\bigg[\frac{1}{R_{x}^{2}} \nonumber \\
 & & \mbox{ \hspace{1em}} -\varepsilon_{\mathrm{dd}}(t) \left(\frac{1}{R_{x}^{2}}+\frac{3}{2}
\frac{f(R_{x}/R_{z})}{R_{x}^{2}-R_{z}^{2} }\right) \bigg] \label{eq:rxmotion} \\
\ddot{R}_{z}& = & - \omega_{z}^{2}(t)R_{z}+\frac{15 g N}{4 \pi m R_{x}^2}
\bigg[\frac{1}{R_{z}^{2}} \nonumber \\
 & & \mbox{ \hspace{1em}} +2\varepsilon_{\mathrm{dd}}(t) \left(\frac{1}{R_{z}^2}+\frac{3}{2}
\frac{f(R_{x}/R_{z})}{R_{x}^{2}-R_{z}^{2} }\right) \bigg].
\label{eq:rzmotion}
\end{eqnarray}
Solving these equations under particular choices of the dependence
of $\omega_{j}(t)$ upon time gives the evolution of the density
and phase of the condensate when undergoing e.g.\ free expansion
(trap switched off), or shape oscillations (time-modulation of the
trap). A time-dependent dipolar coupling is also possible as
explicitly indicated. As mentioned above, our results show that
previous calculations based upon the Gaussian variational ansatz
of the aspect ratio $\kappa(t)$ during ballistic expansion
\cite{giovanazzi03}, or during the parametric excitation of the
quadrupole shape oscillation mode using dipolar interactions as a
time-dependent coupling \cite{giovanazzi2002b}, are exact in the
Thomas-Fermi limit.

\begin{figure}[htbp]
\begin{center}
\centerline{\epsfig{figure=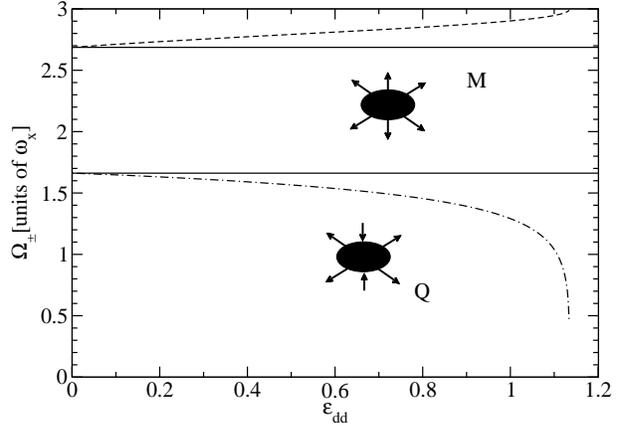,
  width= 7cm,angle=-90}}
\end{center}
\vspace{0ex} \caption{The monopole (dashed) and quadrupole
(dash-dotted) oscillation frequencies in units of the radial
trapping frequency, as functions of $\varepsilon_{\mathrm{dd}}$.
The solid lines are for a BEC with only $s$-wave contact
interactions, $\Omega_{\pm}^{0}$. The trap aspect ratio is set at
$\gamma=240/170$. The frequencies become complex at precisely the
value of $\varepsilon_{\mathrm{dd}}$ at which the equilibrium
solution becomes unstable. However, other instabilities, due to
phonons, are possible whenever $\varepsilon_{\mathrm{dd}}>1$. }
\label{fig:frequencies}
\end{figure}

\begin{figure}[htbp]
\begin{center}
\centerline{\epsfig{figure=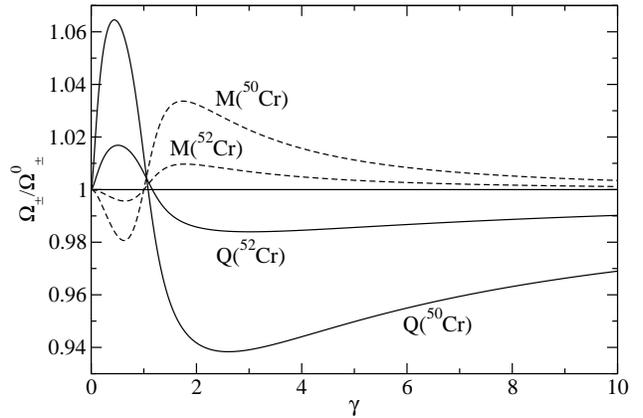,
  width= 7cm,angle=-90}}
\end{center}
\vspace{0ex} \caption{The monopolar (dashed) and quadrupolar
(solid) frequencies as fractions of their $s$-wave-only
frequencies, $\Omega_{\pm}^{0}$, of $^{52}$Cr
($\varepsilon_{\mathrm{dd}}=0.089$) and $^{50}$Cr
($\varepsilon_{\mathrm{dd}}=0.36$), plotted as functions of the
trap anisotropy  $\gamma$.} \label{fig:relativef}
\end{figure}

Linearizing (\ref{eq:rxmotion}--\ref{eq:rzmotion}) around the
equilibrium solution (\ref{eq:Rxsol}) gives the following
frequencies for small amplitude scaling oscillations
\begin{equation}
\Omega_{\pm}^{2}=\frac{1}{2}\left(h_{xx}+h_{zz} \pm
\sqrt{(h_{xx}-h_{zz})^{2}+4h_{xz}h_{zx}} \right)
\end{equation}
where
\begin{eqnarray}
h_{xx} & = &\omega_{x}^{2}+3\omega_{x}^{2}
\frac{1-\varepsilon_{\mathrm{dd}}\left[\frac{1-2\kappa^{2}}{1-\kappa^{2}}+\frac{\kappa^{2}(4\kappa^{2}+1)f(\kappa)}{
2(1-\kappa^{2})^{2}}\right]}{1+\varepsilon_{\mathrm{dd}}\left[\frac{3
\kappa^{2}f(\kappa)}{2(1-\kappa^{2})}-1\right]}
\nonumber \\
h_{zz} & = & \gamma^{2}\omega_{x}^{2}+ 2
\omega_{x}^{2}\kappa^{2}\frac{1+\varepsilon_{\mathrm{dd}}\left[\frac{5-2\kappa^{2}}{1-\kappa^{2}}
-\frac{3(\kappa^{2}+4)f(\kappa)}{2(1-\kappa^{2})^{2}}\right]}{1+\varepsilon_{\mathrm{dd}}\left[\frac{3
\kappa^{2}f(\kappa)}{2(1-\kappa^{2})}-1\right]}
\nonumber \\
h_{zx} & = & 2h_{xz}= 2 \omega_{x}^{2} \kappa
\frac{1-\varepsilon_{\mathrm{dd}}\left[\frac{1+2\kappa^{2}}{1-\kappa^{2}}-\frac{15
\kappa^{2}
f(\kappa)}{2(1-\kappa^{2})^2}\right]}{1+\varepsilon_{\mathrm{dd}}\left[\frac{3
\kappa^{2}f(\kappa)}{2(1-\kappa^{2})}-1\right]} \ .
\end{eqnarray}
$\Omega_{-}$ corresponds to quadrupole, and $\Omega_{+}$ to
monopole (breathing) shape oscillations, respectively
\cite{yi01,goral2002b}, see Fig.\ \ref{fig:frequencies}. Note that
the frequencies do not depend on $g$ directly, only through
$\varepsilon_{\mathrm{dd}}$. Fig.\ \ref{fig:relativef} illustrates
that certain trap anisotropies maximise the effects of the dipolar
interactions upon the collective excitation frequencies.

\emph{Conclusion} --- Dipolar condensates promise to open a new
chapter in the study of controllable superfluids. At first sight,
the non-local and anisotropic form of dipolar interactions also
make the elucidation of the static and excited states of a dipolar
BEC non-trivial, i.e.\ direct numerical calculation is for
instance computationally heavy in comparison to standard BECs with
contact interactions. We have shown, however, that there exist
simple exact solutions which should prove useful for quantitative
analysis of upcoming experiments.

\vskip 5mm

It is a pleasure to thank Gabriel Barton for discussions and
Tilman Pfau and his group for sharing their results with us. We
would like to acknowledge financial support from the UK
Engineering and Physical Sciences Research Council (D.O'D.), the
European Union (S.G.), and the Royal Society (C.E.).

\begin{center}
    \rule{25mm}{0.3mm}
\end{center}

\end{document}